\documentclass[sn-mathphys,Numbered]{sn-jnl}
\usepackage{graphicx}
\usepackage{multirow}
\usepackage{amsmath,amssymb,amsfonts}
\usepackage{amsthm}
\usepackage{mathrsfs}
\usepackage[title]{appendix}
\usepackage{xcolor}
\usepackage{textcomp}
\usepackage{manyfoot}
\usepackage{booktabs}
\usepackage{lmodern}
\usepackage{algorithm}
\usepackage{algorithmicx}
\usepackage{algpseudocode}
\usepackage{listings}
\usepackage{hyperref}
\usepackage{tabularx}
\title{The Quantum Cryptography Approach: Unleashing the Potential of Quantum Key Reconciliation Protocol for Secure Communication}
\author*[1]{\fnm{Neha} \sur{Sharma}}\email{nehasharmablp461@gmail.com}

\author[2]{\fnm{Vikas} \sur{Saxena}}\email{vikas.saxena@jiit.ac.in}

\affil*[1,2]{\orgdiv{CSE and IT}, \orgname{Jaypee Institute of Information Technology}, \orgaddress{\street{Sector 62}, \city{Noida}, \postcode{201309}, \state{Uttar Pradesh}, \country{India}}}
\equalcont{These authors contributed equally to this work.}
\begin{document}
\maketitle
\textbf{Abstract} Quantum cryptography is the study of delivering secret communications across a quantum channel. Recently, Quantum Key Distribution (QKD) has been recognized as the most important breakthrough in quantum cryptography. This process facilitates two distant parties to share secure communications based on physical laws. The BB84 protocol was developed in 1984 and remains the most widely used among BB92, Ekert91, COW, and SARG04 protocols. However the practical security of QKD with imperfect devices have been widely discussed, and there are many ways to guarantee that generated key by QKD still provides unconditional security.
This paper proposed a novel method that allows users to communicate while generating the secure keys as well as securing the transmission without any leakage of the data. In this approach sender will never reveal her basis, hence neither the receiver nor the intruder will get knowledge of the fundamental basis.Further to detect Eve, polynomial interpolation is also used as a key verification technique. In order to fully utilize the quantum computing capabilities provided by IBM quantum computers, the protocol is executed using the Qiskit backend for 45 qubits. This article discusses a plot of \% error against $alpha$ (strength of eavesdropping). As a result, different types of noise have been included, and the success probability of the desired key bits has been determined. Furthermore, the success probability under depolarizing noise is explained for different qubit counts.Last but not least, even when the applied noise is increased to maximum capacity, a 50\% probability of successful key generation is still observed in an experiment. 

\maketitle
\section{Introduction}\label{sec1}
Quantum key distribution uses the fundamental laws of quantum physics \cite{bennett1992quantum}.The objective is to create a shield that cannot be broken, allowing secure communication between distant parties. As part of the first QKD protocol, uncertainty, and non-cloning principles prevent an eavesdropper (Eve) from correctly decrypting shared messages and hiding them from Alice (Sender) and Bob (Receiver). \cite{lo1999unconditional} Coupled with these guarantees, shared messages are considered information-theoretically secure, and quantum key distribution facilitates the exchange of secret keys used for cryptography. An encryption system ensures that confidential information is securely transmitted from one place to another, allowing only the intended recipient to decrypt it. 
Public-key cryptography's security depends on the factorization product of large prime numbers. As a result of technological advances in the computing field and the introduction of novel mathematical methods, the problem of factoring integer products has become less complicated \cite{slayton2022democratizing}. As an example of an asymmetric algorithm, RSA (Rivest, Shamir, and Adleman) decomposes large integers into prime components. To find the many prime factors of the multiplication result requires a lot of effort and time. A factorial polynomial-time method based on quantum computers is Shor's algorithm \cite{monz2016realization}.
Cryptography techniques used by the public have become more vulnerable since 1998\cite{steane1998quantum}. As a result, quantum computing technology is gaining exponential interest. Scientists have demonstrated that quantum computers defeated RSA and other contemporary encryption methods in polynomial time. Due to the increasing use of quantum computing approaches, new cryptography methods are being developed. Using quantum physics concepts, quantum cryptography is a new field of study. Since the 1980s, quantum computing has been based on concepts from quantum mechanics, including entanglement and superposition.Computing became possible after 1984 with the Quantum Computer.As the first result of the quantum computer, 16 qubits were used to implement the Sudoku game.The first quantum plan of IBM was revealed in 2020 \cite{pal2020solving}.
There are 127-qubit processors, quantum circuits that cannot be accurately emulated in conventional computers, and a new architecture from IBM, Eagle, that is designed for processors with more and more qubits.
In addition, compared to a previous experiment in 2017, IBM has given a 120x speedup within the capacity to imitate a particle utilizing Qiskit Run Time, IBM's containerized quantum computing benefit, and its programming dialect. Additionally, IBM has achieved 120x speed increases in the ability to simulate a particle using its containerized quantum computing service, Qiskit Run Time, and programming language. To meet the previously stated milestones on IBM's road map, IBM plans to dispatch its 433-qubit processor later in 2023, IBM Osprey. However over the conventional computer, Quantum key distribution is now a popular research area, with new protocol proposals and evidence of their security gained \cite{huttner1995quantum}. In order to improve these protocols' effectiveness, privacy, and compatibility, \cite{pirandola2015high} numerous studies have already been carried out. In order to improve key rates, reduce quantum bit error rates, and expand the range of secure communication, scientists have experimented with a variety of methods \cite{wang2014field}.
Scientists, however, have primarily focused on identifying and correcting security vulnerabilities in existing quantum key distribution systems \cite{browne2017quantum}, instead of designing novel protocols. Multiple QKD systems have been compromised by attacks such as photon-number splitting and detector blinding. Due to these vulnerabilities, several strategies have been developed and implemented, including the decoy-state method and secure detectors \cite{zhao2008quantum}. 
The integration of quantum key distribution (QKD) into existing communication networks is an essential aspect of quantum cryptography research\cite{hwang2003quantum}. Integrating QKD into optical networks is one solution. Today's communication frameworks are based on optical networks. Several studies \cite{bernstein2012security}have been conducted to assess the feasibility of deploying QKD within wavelength-division multiplexing (WDM) networks and passive optical networks (PONs). According to these studies, QKD can enhance the security of optical networks without compromising their efficiency. Furthermore, the integration of QKD into optical networks has led to novel service models, such as Key-as-a-Service (KaaS). As part of KaaS, QKD is seamlessly incorporated into the underlying optical infrastructure, facilitating secure key distribution within virtual optical networks (VONs). \\
The security of continuous variable QKD and other protocols must be improved to defend against future attacks. Furthermore, the practical deployment of quantum cryptography systems, including factors such as size reduction, cost-effectiveness, and harmonic integration with existing infrastructure, remains a critical focus of ongoing research{kumar2015coexistence}. It is possible to view the sharing of confidential information between multiple parties as a high-risk activity. Data-sharing infrastructures must be safe for researchers, administrators, foundations, and the general public.
For information to be transmitted securely by many parties, it must be encrypted with a unique key that can only be identified by the sender (Alice) and recipient (Bob). An encrypted key is generated by the sender and sent to the receiver using various communication methods. Even though quantum key distribution (QKD) offers absolute data security and is currently used in commercial applications /cite[huth2016information], it must overcome several significant obstacles such as secret key rate, distance, size, cost, and practical security in order to become a widely adopted technology. As a result of the work presented in this paper, an important technique for key verification without leaking any information has been developed.\\
$\bullet$ The main difference between the modified and the existing BB84 protocol comes after key generation. Instead of both parties revealing their key bits, only Alice knows the correct encoded basis.\\
$\bullet$ Alice and Bob at first calculate the errors in their shared secret key (key sifting). The two parties remove the wrong bits from the key and the presence of Eve is verified as a result.\\
$\bullet$The detection of Eve and the key verification do not leak information. Therefore, privacy amplification is not necessary in post processing.\\
There is a limit in the amount of key bits that can be used by an intruder when it comes to utilizing shared information, which is governed by the nature of the QkD protocol \cite{bruss1998optimal}\cite{liu2017novel}.\\  
Here, key verification algorithms are redesigned to overcome the original paradigm \cite{berlekamp1970factoring}. In the proposed work, no key information is compromised in order to remove incorrect key bits. However, in CASCADE error correction algorithm, a few key bits are revealed, thus resulting in shorter key bits.
A section 2 of this article discusses the background of this article and related work to provide a deeper understanding of the work. The enhanced QKD is demonstrated in section 3 from basic to advanced. In Section 4, the methodology for the proposed study is described. A key verification and detection of Eve with scenarios is incorporated in Section 5. A discussion of simulation and experimental results follows in section 6. A conclusion and future projects are also presented to provide researchers with a deeper understanding of Section 7.
\section{ Related Work}\label{sec2}
 A comparison of various approaches is presented to provide an overview of the work done on keys reconciliations based on QKD protocol between 2016 and 2023. According to the authors, their method reduces computational costs and communication delays compared to other one-way information reconciliation schemes. Slepian-Wolf's limit constrains existing reconciliation schemes due to the presence of information leakage\cite{kiktenko2017symmetric}. The reconciliation process does not guarantee that the secret keys are identical after the reconciliation process, so there is the potential for errors to propagate \cite{lee2018improved}.\\
 Based on Table 1, the source, observation, gaps, and a few other parameters are analyzed. By utilizing the QKD protocol, the author has presented a number of new applications and approaches. Using their approach, information leakage is mitigated and identical secret keys are generated, so the chance of error propagation is minimized. Based on the authors' assessment, their method offers superior computation efficiency and reduced communication delays. For the exchange of keys, \cite{martinez2022modification} proposes the NewHope algorithm. In this algorithm, the limitations are the length of the shared secret, the distance covered, and the dimension of the calculation. As per pre-coding scheme (DVB-S2), it is based on digital video broadcasting. CV-QKD has a lot of work to do when it comes to improving the distance, security, and reliability of the code. A second proposal on Newton's polynomial interpolation is given by \cite{kurt2020polynomial}.
It is necessary to reduce computational complexity and in future work to optimize block lengths and numbers of blocks in the protocol. 
\begin{table*}[!htp]
\small
\label{pfperformance}
\caption{\\Comparison of the Existing QKD protocols used in various studies}
\centering 
\resizebox{12cm}{8.5cm}{
\small \begin{tabularx}{1.5\textwidth}
{  
 | >{\raggedright\arraybackslash}X 
 | >{\raggedright\arraybackslash}X
 | >{\raggedright\arraybackslash}X
 | >{\raggedright\arraybackslash}X 
 | >{\raggedright\arraybackslash}X 
 | >{\raggedright\arraybackslash}X
 | >{\raggedright\arraybackslash}X 
 | 
}
\hline
 Source & Objective & Key Reconciliation Scheme & Correction Code & Efficient Approach & Gaps & QKD \\ 
\hline
Guo et al.\cite{guo2020comprehensive}&Sliced error correction on a heterogeneous computing structure&$\checkmark$&$\checkmark$& $\checkmark$&In future pipeline structure must be designed to improve the throughput&$\checkmark$ \\
\hline
Schimpf et al.\cite{schimpf2021entanglement}&Entanglement-based quantum key distribution (QKD)&$\checkmark$&$\times$ &$\checkmark$&QBER during QKD can be optimized by mild time filtering&$\checkmark$ \\
\hline
Dirks et al.\cite{dirks2021geoqkd}&Implemented an untrusted and trusted mode BBM92 QKD protocol&$\checkmark$&$\checkmark$& $\checkmark$&In future GEO QKD implementation must be considered as a serious addition to the design of a space-based quantum communication network&$\checkmark$ \\
\hline
Borisov et al.\cite{borisov2022asymmetric}&Compared the performance of several asymmetric and symmetric error correction schemes using a real industrial QKD setup&$\checkmark$&$\times$&$\checkmark$&Improvement in symmetric and new asymmetric schemes demonstrated close efficiencies and average numbers of decoding iterations&$\checkmark$ \\
\hline
Tseng et al.\cite{fang2022improved}&Sampling-based approaches& $\checkmark$&$\checkmark$&$\checkmark$&This would require extending the quantum sampling technique to support such measurements&$\checkmark$ \\
\hline
Wang et al.\cite{wang2023non}&Non-Gaussian reconciliation method&$\checkmark$&$\times$&$\checkmark$&The layered belief propagation decoding algorithm is introduced for error correction, which can greatly reduce the decoding complexity and improve the decoding speed&$\checkmark$ \\
\hline
Lin et al.\cite{lin2023counterfactual}&Security of counterfactual QKD in an untrusted detector scenario&$\checkmark$&$\times$ & $\checkmark$&The security of CQKD seems to be untrusted detectors; it is assumed to be protected from a well-characterized state preparation process&$\checkmark$ 
 \\
 \hline
\end{tabularx}
}
\end{table*}
As part of a 511 km fibre trunk connecting two remote metropolitan regions without the need for a reliable repeater, the author implemented a \cite{chen2021twin} twin-field quantum key distribution (QKD10). In this way, long-distance fibre quantum networks can transfer quantum states efficiently and propagate single photons with stability. The author proposes \cite{lucamarini2018overcoming} in this paper an approach in which QKD has a rate-distance limit of 550 km, but the proposed system has quantum repeaters in it, so it can handle a decent amount of noise even at 550 km on regular optical fibre. In addition to making quantum communications even safer, it's a huge step forward.
A novel QKD system time is proposed in \cite{wang2022twin} to handle over 140 dB channel loss and reach 833.8km, a record for QKD on fibre. As a result of the four-phase double-field protocol and good setup, the secure key rate over similar distances is increased by over two orders of magnitude. In terms of reliability and efficiency over 1,000 km, this is a major step forward.The author proposes \cite{gu2022experimental} a fault profile that can be compared to experiments in this work. A finite key analysis is also suggested as a means of preventing coherent attacks. They test the feasibility of our protocol with an experimental implementation. The proposed protocol was also tested and obtained 253 bits per second at a channel loss of 20dB. It also significantly increases key rate and transmission distance when compared to previous QC methods. It shows the practicality of using secure QC to address device defects.
Based on the results of the proposed  protocol, it has \cite{li2023breaking} been proven that it can beat the rate-distance limits across networks with at least ten communication partners. The author increased the speed of the secret key by more than two times and extended the transmission distance. Also, they assess the safety of the protocol in a composable framework, taking into account participant attacks, in a finite-size system.
This study  \cite{zhou2023device} proposes an extremely efficient, realistic, single-photon source protocol (DI-QSDC). Entanglement channel is constructed by the communication parties using single photons in a well-known manner. There is no relationship between the message leakage rate and the loss of photons. Its secure communication distance is approximately six times that of the original protocol and its practical communication efficiency is approximately 600 times that of it.
The author \cite{yin2107experimental} has developed a QDS system which uses secret sharing of quantum keys, universal hashing, and a single pad to sign documents. The proposed system requires a  256-bit key with a security limit of 1019 characters. One megabit document is signed more than 108 times more efficiently than a previous QDS protocol.

\section{Design and Methodology of Proposed Protocol} A variety of cryptographic applications can benefit from the proposed method to generate safe secret keys for both the sender and receiver. BB84 protocol pseudo code is described below with its basic structure. The first step Alice takes is to randomly generate her bits, encode them on the random basis as in BB84 protocol, and transmit them over fiber optics to Bob. The proposed protocol is performed on the IBMQ platform, along with the quantum circuit. As a result, we prepare a circuit in which each qubit corresponds to each photon in the circuit. A quantum gate transforms the qubit state to encode bits and bases in the polarization of the photon. According to Bob, a quantum circuit is measured by randomly choosing the basis for measurement and then applying the appropriate quantum gates. The measurement results in each party receiving a measurement bit and thus forming an initial key.\\
In original BB84 in order to generate the key bits both parties publicly announce their encoding and decoding basis. Decoding and encoding outputs that correspond to different bases of encoding and decoding are discarded. Using the common bases both parties use, they compare a small set of measurement bits randomly to detect Eve's presence. It is only when Bob publicly announces his measurement basis that the novelty of the proposed work is seen instead of this practice. Only Alice knows encoding basis, and she passes commands to Bob to generate sets (as explained in section 4.1).Through set generation, Alice and Bob get $M*m$ rows and columns of key bits. And then by using polynomial interpolation they check the presence of Eve(explained in section 5).  
Once the absence of Eve is confirmed the raw key bits can be pass for key reconciliation. In our work, we have also proposed a key reconciliation approach that uses polynomial interpolation to detect and remove the wrong key bits, just as we did during Eve detection.Since we had already detected and removed errors from the key bits while checking for Eve, this step is not mandatory.Error-correction techniques, however, should always be used to ensure that no bits are wrong. \\ 
\textbf{Steps of Pseudo Algorithm} \\
1. Alice randomly generates 2 binary sequences of same length $\left(4^* \mathrm{N}\right)$ namely Alice\_bits and Alice\_basis (where $\mathrm{n} \in \mathrm{Z}$ )\\
2. Alice creates a quantum circuit \\
qc= QuantumCircuit $\left(4^* {n}\right)$,$\left(4^* {N}\right)$\\
3. for $i$ in {range}$\left(4^* {n}\right)$ :
\begin{lstlisting}
       if Alice_basis[i]==0: 
               if Alice_bits[i]==0:
                        qc.i(i)
                else:
                        qc.x(i)
        else:
                if Alice_bits[i]=0: 
                         qc.h(i)
                else: 
                         qc.x(i)
                         qc.h(i)
\end{lstlisting}
end for loop \\
4. Bob also randomly generate his binary sequence of bases of same length called bob\_basis and measure the qc on the basis of bob\_basis \\
5. for $i$ in {range}$\left(4^* {n}\right)$ :
\begin{lstlisting}
             if bob_basis[i]==0 :
                      qc.measure(i,i)  
             else:
                      qc.h(i) 
                      qc.measure(i,i)
\end{lstlisting}
6. Bob then on classical channel reveals his bases and Alice don't(now only Alice knows the common bases of measurement) \\
7. Alice commands Bob to shuffle his measurement output for the set generation such that $\mathrm{M}^* \mathrm{~m}=4^* \mathrm{N}$ (where $\mathrm{M}, \mathrm{m}$ are the no. of rows and columns respectively) \\
8. Each row has $m$ no. of bits divided in s chunks such that $\mathrm{n} * \mathrm{~s}=\mathrm{m}$ (where $\mathrm{n}$ is no. of bits in each chunk) both the bases.\\
9.All the 4*n bits(of both parties) are divided in the rows and columns on the demand of Alice in such a way that each row contains either all the elements corresponding to common bases or to different bases but not to both the basis.\\
10. All the s chunks are then grouped together and are converted into decimal numbers.\\
11. Alice then randomly generates another sequence corresponding to each decimal converted rows(of the length of converted row) called alice\_y \\
12. Using his decimal no. as ${x}$ coordinated and the corresponding values of alice\_y as $y$ coordinated Alice gets ${s}$ no. of $\left(x_i, y_i\right)$ pairs. coefficients.\\
13. Alice then generates a unique $s-1$ degree degree polynomial(using polynomial interpolation or langrange interpolation or any other technique) passing through those $s (x_i , y_i)$ coordinates and stores their coefficients. \\
14. Alice sends the Alice\_y to Bob and Bob also tries to generate the polynomial(by same technique) using his decimals no. as his x-coordinates and corresponding element of Alice\_y as y coordinate and stores the coefficients.\\
15. Number of removed element $=0$ \\
16. While(no. of removed element $<={s}$):\\
\text{if all the coefficients of Alice and Bob are same}:
$$\Rightarrow x^A{ }_i=x^B{ }_i \text {(for all }\mathrm{i} \text { of the same row)}$$
else:\\
Alice and Bob randomly removes an element and and try to generate a polynomial of $s-j$ degree passing through the remaining $s-j-1$ points and compare their coefficients.\\
\text{if all the coefficients of Alice and Bob are same}:\\
      \text {return no. of removed elements along with their indices and values}.
\\
else:\\
remove another element and continue(repeat the step of generation of polynomial)
While loop end  \\
17. Finally Alice calculate E$_i$(bit error) corresponding to every row on the basis of no. of removed element and then calculate average bit error rate $\epsilon = \sum E_i$ over only the common basis. \\
18. if $\varepsilon<\boldsymbol{\tau}$ :\\
continue the QKD protocol and pass the shifted key bits for post processing \\
else:\\
restart the QKD protocol again
If $\epsilon$ $<$ $\tau$ then we convert the decimal numbers in to binary numbers again and generate a set of rows and columns as earlier and then each chunk of newly generated row is converted again in to decimal no.
The process of removing the error performed again using polynomial interpolation as before. This very second stage of error detection is optional but it is a good practise in the QKD protocol. 
All the parameters like $m$, $M$, $s$, $r$ etc are adjusted accordingly, as the size of remaining key bits is same as earlier.
After the process of key reconciliation using polynomial interpolation if the key bits has become very small then we pass the key for privacy amplification or else use it as key in the encryption techniques.  
Figure 1 is depicting the methodology for the generation of authentic key. The protocol begins by defining the parameters that will be utilized in the following phase. Alice is going to use a serial-to-parallel converter to split the organized key into $M*m$ rows and columns.\\
\begin{figure}
\centering
\includegraphics[width=0.7\textwidth,height=7.5cm]{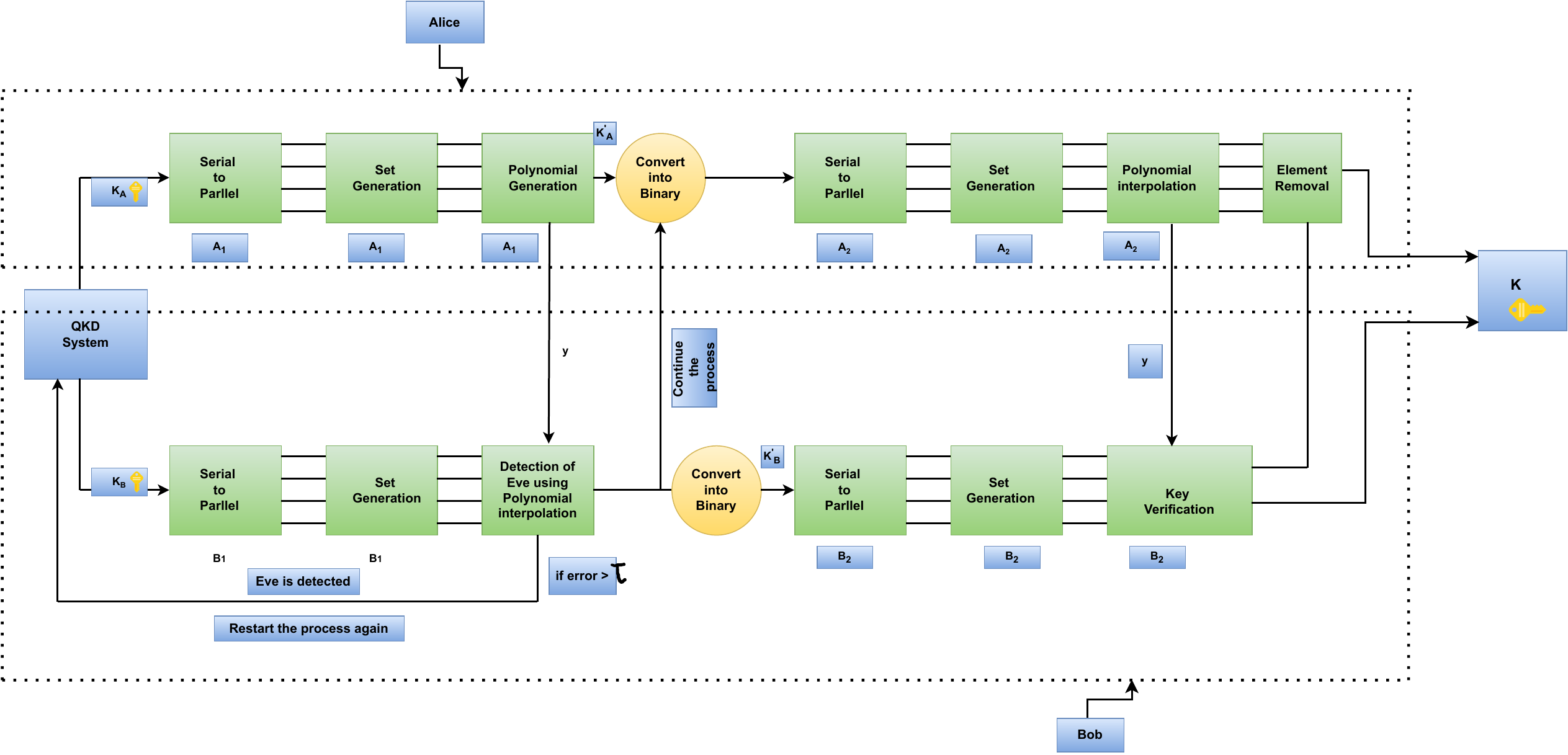}
\caption{The proposed methodology is an appropriate technique for both the sender and receiver sides to generate safe secret keys}
\end{figure}
\section{Key Verification with Polynomial Interpolation}
For the verification of the generated key, once the enhanced $BB84$ protocol is implemented, Alice and Bob will have a key in the form of a bit string. With some relief information for Eve and some erroneous bits.
Here, $\epsilon$ is donated as QBER (Quantum Bit Error Rate)and the maximum tolerable error is represented as $\tau$. 
\newline
If $\epsilon$ $<$ $\tau$ then raw key bits will further send for key verification else we restart the protocol and generate the raw key bits. Table 2 is representing all the notation used in the study.
\begin{table}
\caption{Representation of Notations used in the proposed protocol}
\centering
\begin{tabular}{llllllllllllllll}
\hline\noalign{\smallskip}
$S.No.$ & Symbol & Meaning  \\
\hline
1 & $B_j^A$ & Bit string of $j^{th}$ block \\
2 & $C_{ij}^A$ & Bit string in chunks \\
3 & $C$ & Chunk \\
4 & $M$ & Total number of elements in a column after set generation \\
5 & $q$ & Prime number \\
6 & $p(x)$ & Coefficient of Polynomial \\
7 & $x_0, y_0$ & Data points \\
8 & $k$ & Key (bit string) \\
9 & $j$ & Decimal representation of the corresponding chunk \\
10 & $s$ & Number of chunks, where every chunk contains several bits\\
11 & $\epsilon$ & Quantum Bit Error Rate \\
12 & $\tau$ & Maximum Tolerable Rate \\
13 & $\alpha$ & Strength of eavesdropping \\
14 & $m$ & Total number of bit in a row  after set generation\\
15 & $n$ & Number of bits in each chunks after set generation \\
16 & $c$ & Total number of common base \\
\noalign{\smallskip}
\hline
\end{tabular}
\end{table}
In the next steps by explaining three sections complete concept of protocol is described \\
1. Set Generation \\
2. Polynomial Generation  \\
3. Key Verification  \\
\subsection{Set Generation}
First Alice divides the bit string into $M$ blocks, where $B_j^A$ denotes the Bit string of the $j^{th}$ block $j$ $\in$ $(0,1,....M-1)$ and each block has $m$ bits thus. 
\begin{equation}
M\cdot m=4\cdot N
\end{equation}
Alice split each block’s bit string in chunks by $C_{ij}^A$.\\
Each chunk has $n$ bits and $i$ $\in$ $(0,1,......s-1)$.
Each block’s bit string got divided into $s$ no. of chunks where every chunk contains several bits. 
\begin{equation}
n.s=m
\end{equation}
$s$ is selected in such a way that every chunk is unique in each block it should also be less than $2^n$. 
We choose the size of the chunk and block such that it becomes infeasible to have two chunks with the same bit string. After that, each chunk is converted to its decimal number from the binary. 
 \begin{equation}
C_{i,j}^A \rightarrow a_{0,j},a_{1,j}.....,a_{s-1,j} 
\end{equation}
$i$ $\in$ $(0,1,......s-1)$ \\
$C$ = Chunk $j$ = decimal representation of corresponding chunks Such that
$a_g,_j$ $\neq$ $a_h,_j$ if $g$ $\neq$ $h$ \\
For $0$ $\le$ $g$,$h$ $\le$ $s-1$\\
This simply means that after converting every chunk to its decimal number, the two chunks should not have the same decimal number. The bit string of decimal numbers should contain all distinct numbers.  
\begin{equation}
s_j^A = a_{0,j}, a_{1,j}, a_{2,j}, \ldots, a_{s-2,j}, a_{s-1,j}
\end{equation}
$j$ = $0,1,....M-1$\\
A string that contains all the decimal numbers. Below Figure 2 is a pictorial representation of steps to set generation for key verification.
\begin{figure}[htp]
\centering
\includegraphics[width=0.7\textwidth,height=4.5cm]{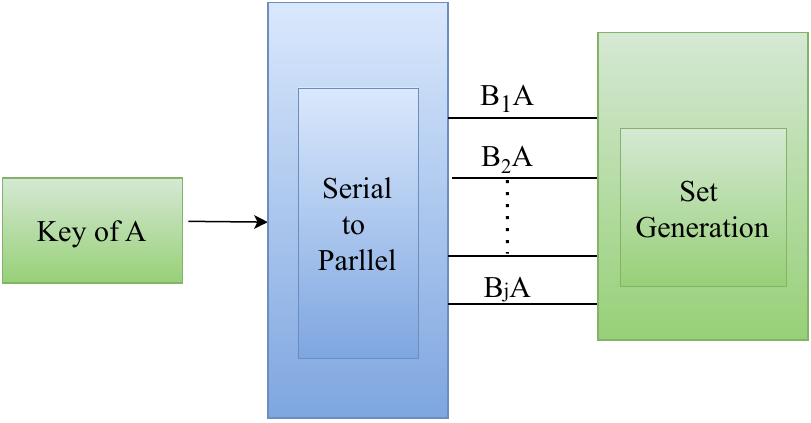}
\caption{Set Generation for key verification}
\end{figure}
\subsection{Polynomial Generation} 
During the generation of each row, Alice randomly generates s distinct decimal numbers. For each row, the decimal numbers are used as x coordinates, and the randomly generated number corresponding to them is used as y coordinates. By using these (x, y) coordinates, a unique polynomial passing through these points can be derived. As shown in theorem 1, there exists a polynomial of degree s-1 that passes through s distinct points. It is possible for Alice to perform polynomial interpolation, Lagrange's interpolation, Newton's polynomial interpolation, etc. In order to obtain that unique polynomial of degree $s-1$, passed through $s$ $(x_i, y_i)$ coordinates, where $i$ is $/in$ $(0,1,......s-1)$.
Bob receives the randomly generated sequence used as $y$ coordinates from Alice and stores the coefficients of the derived polynomial. A chunk of decimal numbers sent by Bob becomes x coordinates, and the sequence sent by Alice becomes y coordinates. Bob attempts to mimic Alice's key bits. So if Alice and Bob have the same decimal numbers, then they will both derive the same polynomial, as indicated by their coefficients. The coefficients that do not match indicate Bob has at least one different key bit, resulting in at least one different decimal number. Then Alice and Bob derive a s-2 degree polynomial at their respective ends and check their coefficients by randomly excluding one $(x_i, y_i)$ coordinate. All the coefficients must be the same to mean they derived the same polynomial. Otherwise, they use all the different combinations until they come up with the same polynomial without including another random coordinate. If they do not obtain the same polynomial after attempting all possible combinations, they separate 2 coordinates and repeat this step until either all coordinates are not included or they have derived the same polynomial. 
\subsubsection{Theorem}
Let $(x_0$, $y_0$, $x_1$, $y_1$, \ldots $x_n$, $y_n)$ be the Euclidean points through which the graph of $g(x)$ passes then there exists a unique polynomial $p(x)$ of degree less than or equal to $n$ such that p$(x_i)$ = g$(x_i)$ = $y_i$ where $i$ $\in$ ${0, 1, 2, \ldots n}$
\newline
To put it another way, there exists a unique polynomial of degree $n$ that passes through $n+1$ points \cite{ridley2021quantum} \\
Here equation can be find that satisfies the above condition by using several interpolating techniques like divided difference, Lagrange interpolation, etc.
\newline
\subsubsection{Proposition}
Let $p(x)$ be a polynomial of degree $n$ over the real field $R$. 
Then the probability that some arbitrary point $(x_i, y_i)$ will lie on the graph of $p(x)$ is negligible. \\
In other words, if we choose a random point without knowing the precise polynomial $p(x)$, the probability that the point will pass through $p(x)$ is negligible. Authors \cite{aji2021survey} have used this proposition to generate a unique polynomial as explained in the theorem to secretly discard the corrupted bits such that Alice and Bob have a secure and correct key.\\
The secrecy of the key bits depends on the secrecy of the polynomial \cite{sharma2021emerging} which is guaranteed by the above proposition \cite{sharma2022efficient}. 
\section{Detection of Eve Using Polynomial Interpolation and Error Correction}
As already discussed in section 4.1 after the process of set generation, all the 4*N qubits are divided into rows and columns. They are assembled in such a way that all the elements of a row belong to either common bases or to different bases but not to both bases. Hence a complete row will contain either all the elements from common bases or all with different bases. The arrangement requires the elements of common and different bases in a way that they can be divided into a rectangular matrix. If the condition is not satisfied then a few bits are discarded such that the remaining bits can form a rectangular matrix. 
\begin{figure}[htp]
\centering
\includegraphics[width=1.2\textwidth,height=6.5cm]{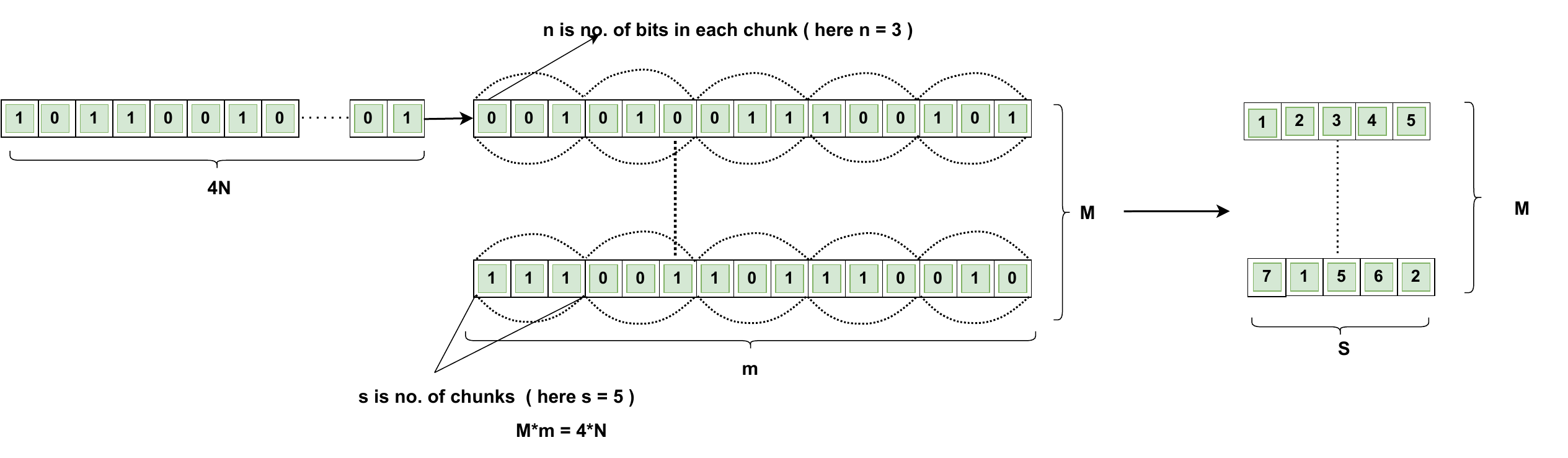}
\caption{Arrangement of $4*N$ bits in $M$ rows and m columns after set generation where each row has $s$ no. of chunks each containing $n$ bits}
\end{figure}
As shown in Figure 3 after set generation, we divide the m bits of each row into several chunks. Each chunk contains n bits and the total s such chunks are there in each row. Each chunk of the bits is converted into decimal numbers. Hence the M*m binary matrix becomes M*s matrix of decimal numbers. $s$ is the number of decimal numbers in each row(here s =5 ).
$$M*m = 4*N$$
$$n*s = m$$
Alice randomly generates a sequence of length $s$ of distinct decimal numbers, corresponding to each decimal converted number of key bits. Alice then treats the decimal numbers of each row as $x$ coordinates and randomly generate $y$ coordinate. \\
\textbf{Alice\_x = (1 2 3 4 5)}
\textbf{Alice\_y = (10 11 12 13 14)}\\
By using polynomial interpolation or Lagrange interpolation Alice will drive an equation that satisfies all the above points. 
\begin{figure}[htp]
\centering
\includegraphics[width=0.7\textwidth,height=3.5cm]{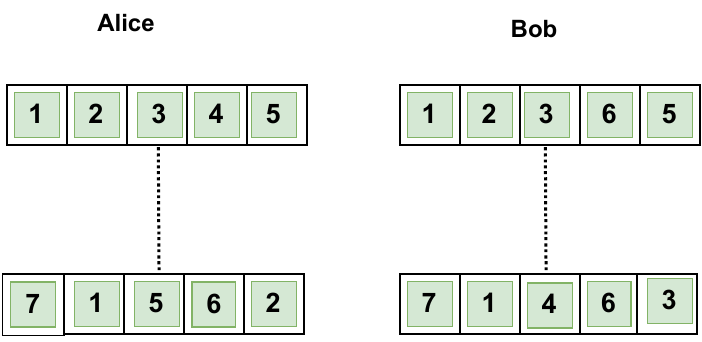}
\caption{Random selection of indices for the equation generation}
\end{figure}
Basically in this case we had ($x_1$,$y_1$), ($x_2$,$y_2$).........($x_5$,$y_5$)
i.e. (1,10),(2,11)...........(5,14) five different coordinates and there is a unique 4-degree polynomial passing through these five points.\\
\textbf{Alice Side:}After driving the equation Alice keeps the coefficient of the polynomial with her. She only sends the Alice\_y to Bob which she had generated randomly and used as the $y$-coordinates in deriving the polynomial .\\
\textbf{Bob Side:}Bob also does the same thing as Alice. He tries to derive the same polynomial using his decimal number as $x$ coordinates and numbers sent by Alice as $y$ coordinates.\\
As also shown in Figure 4 one number of Bob is different in the first row than that of Alice \textbf{(1,10),(2,11).......(3,15),(5,16)} due to which Bob is going to drive different polynomial. Now, Bob will store the coefficient of his polynomial. Alice and Bob will communicate on a classical channel and compare their coefficients. If all the coefficient matches then it means all the ($x_i$,$y_i$) were same for the both Alice and Bob for that particular row. Since we know all the $y_i$ are also the same for both Alice and Bob. Therefore every $x_i$ must also be the same for both parties. Now in case there is any mismatch in the comparison of coefficients it means that at least one of the bits is different for both parties. Hence the decimal number is also different, and at both ends of Alice and Bob, a different equation is derived. To correct the equation Alice and Bob randomly dis-include one of the coordinates and try to drive again a unique polynomial of 3rd degree passing through all considered points.
(In general, derive a polynomial of degree (s-j-1) passing through (s-j) points). Now if they derive the same equation (checked after comparison of coefficient) then as an output they get the removed elements and their index. 

Alice and Bob keep randomly dis-including one coordinate and derive the same polynomial. If after trying all the possible combinations they randomly dis-include two coordinates and continue to repeat the above process till the time they get the same polynomial. As in our case, both the parties will reach to same polynomial after removing just one element i.e., the element at index number 3. Both the parties will derive the same polynomial using the coordinates (1,10),(2,11), (3, 12), and (5,14).
Once they both either reach the same polynomial or dis-include all coordinates Alice maintains the register of dis-included x coordinates, their index number. Then they communicate over the classical channel and calculate the number of bits that were different in the binary representation of the dis-included elements. 
Alice calculate $E_i$ where $i$(1,2,3......M)
Where $M$ is the number of rows in set generation and $E_i$ is the bit error in each row. 
\begin{equation}
\frac{E_i}{m}    
\end{equation}
Where Equation 5 shows the bit error rate.
Finally, Alice calculate the average bit error rate corresponding to the error which belongs to common bases only.
Average bit error rate
($\epsilon$) = $\sum_\frac{E_i}{c}$ 
\newline
$\forall$  i  $\epsilon$ (index of common basis)
\begin{align}
\frac{{E_1 + E_2........E_n}}{{c}} =  \epsilon \leq (\tau)
\end{align}
If the inequality in equation 6 is followed then we can use the keys for the post-processing or else if $\epsilon$ $>$ $\tau$  restart the proposed enhanced QKD Protocol.
If $E_i$ is very large for a row then it should correspond to a row of different basis.
But if Alice selects a row with the correct Basis and still \textbf{$E_i$ for that row is larger than usual, then it can be due to the presence of Eve whose presence corrupted those $m$ bits more than usual bit flips that occur due to imperfect devices.}. For extra security, if any row has $E_i$ larger than usual then that particular row can also be removed even if the $\epsilon \leq (\tau)$
\section{Simulation and Experimental Results}
In this section, we are going to explain the execution of the protocol, results, and observations in several parts. We have executed the protocol on one of IBM's Qiskit backend simulators called Qasmsimulator. For the raw key bits generation, the protocol is simulated for 7 qubits.
\subsection{Execution of quantum circuit and generation of raw key bits} As shown in Figure 5 Alice randomly generates her bits and basis which are then encoded in the quantum circuit using quantum gates. X basis is the rectilinear basis and Z basis is the Hadamard basis, hence before encoding or measuring qubits in Z bases we apply the Hadamard gate and then perform the operation. After generating the bits and bases Alice creates a quantum circuit of 4*N qubits and encodes each bit in each respective basis in each qubit. 
\begin{figure}[htp]
\centering
\includegraphics[width=1\textwidth,height=5.55cm]{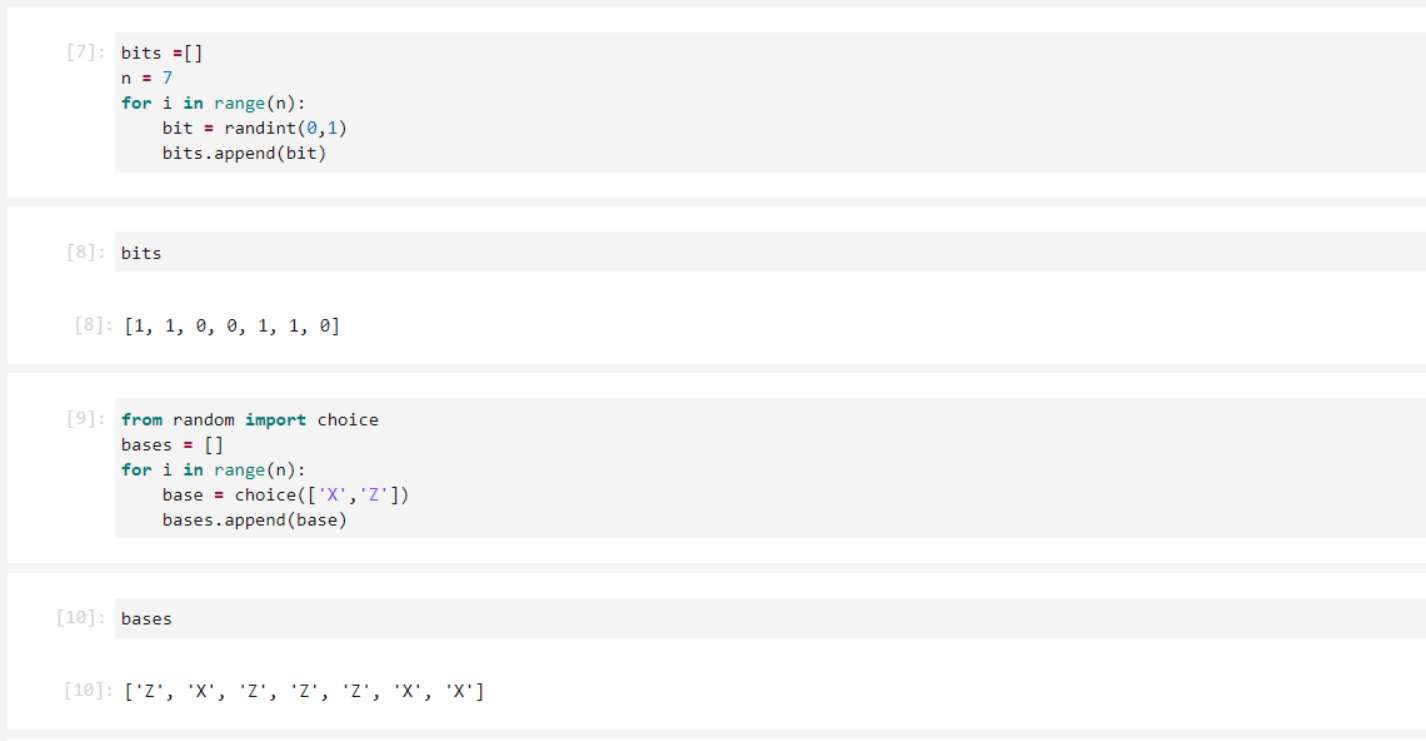}
\caption{Alice randomly generating her bits and her encoding bases}
\end{figure}
Similar to Alice Bob also randomly chooses his basis of measurement and performs the measurement. Figure 6 providing the random generation of Bob's bases and  measurement on the quantum circuit created by Alice. \textbf{The result of the measurement will be the bob's received bit}.\\
The quantum circuit in which Alice encoded her bits and Bob measurement is shown below in Figure 7. Different qubits and different gates are applied before measurement, this is done to encode different bits in different bases. 
\begin{figure}[htp]
\centering
\includegraphics[width=1\textwidth,height=5.55cm]{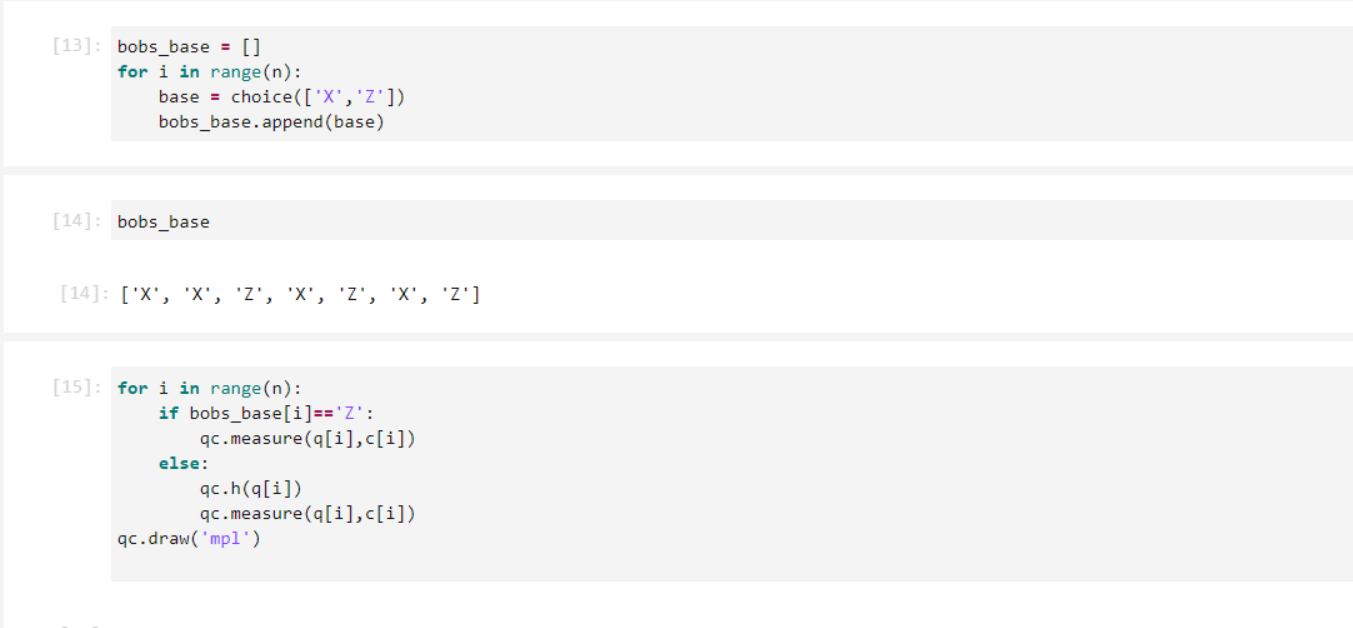}
\caption{Bob randomly chooses his measurement bases and performs measurement on the quantum circuit created by Alice for QKD}
\end{figure}
\begin{figure}[htp]
\centering
\includegraphics[width=1\textwidth,height=5.33cm]{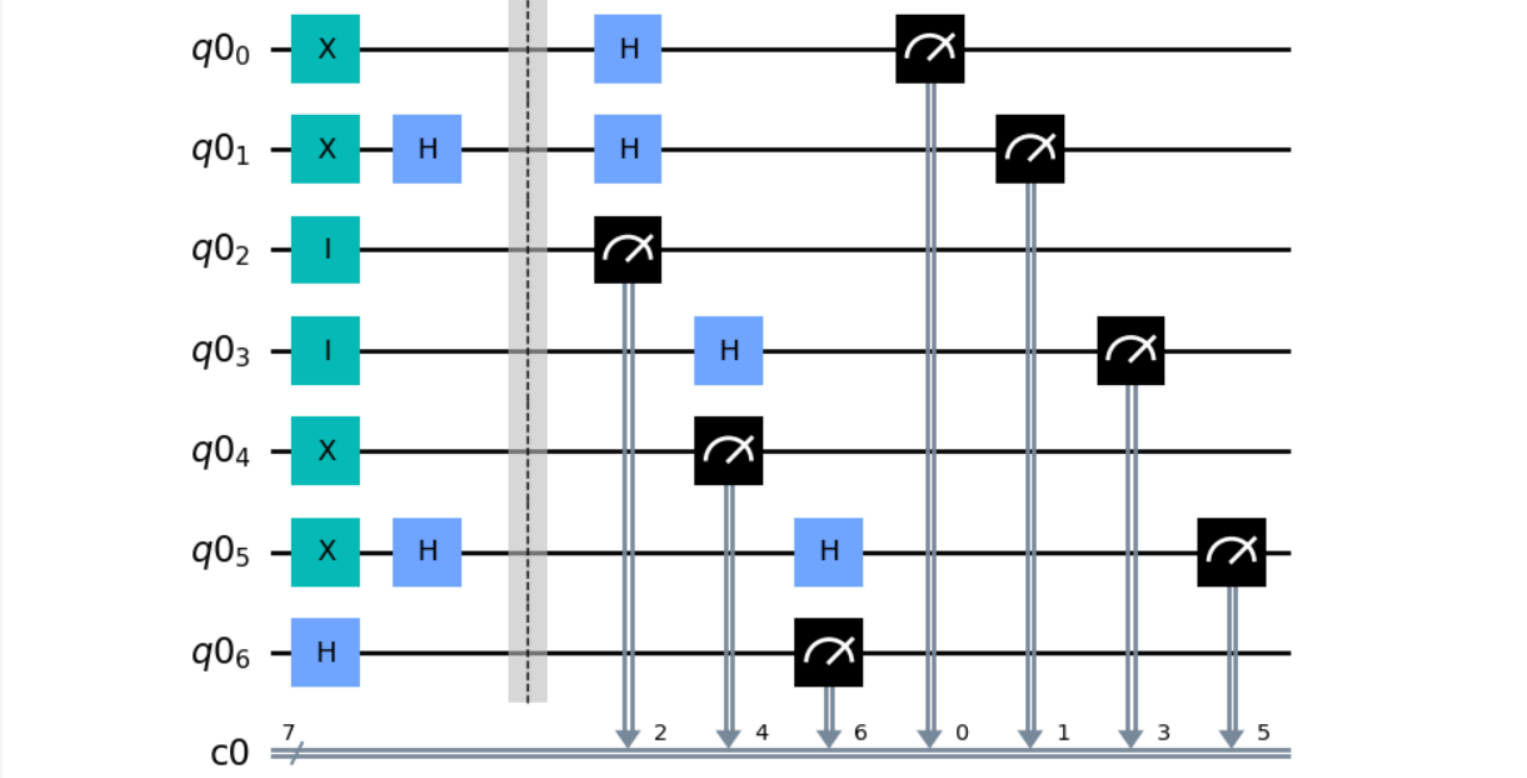}
\caption{7- qubits quantum circuit of the QKD protocol}
\end{figure}
\textbf{Alice's bits}\\
$[1, 1, 0, 0, 1, 1, 0]$\\
\textbf{Alice's bases}\\
$['Z', 'X', 'Z', 'Z', 'Z', 'X', 'X']$\\
\textbf{Bob's' bases}\\
$['X', 'X', 'Z', 'X', 'Z', 'X', 'Z']$\\
\textbf{Common bases}\\
$['X', 'Z', 'Z', 'X']$\\
\textbf{Common bases index}\\
$[1, 2, 4, 5]$\\
The above-shown quantum circuit is executed for 1024 shots on the Qiskit backend and the measurement result (Bob's measurement bits) is plotted. In Figure 8 Bob's measurement bits along with their counts have been plotted.
\begin{figure}[h]
\centering
\includegraphics[width=1\textwidth,height=5.5cm]{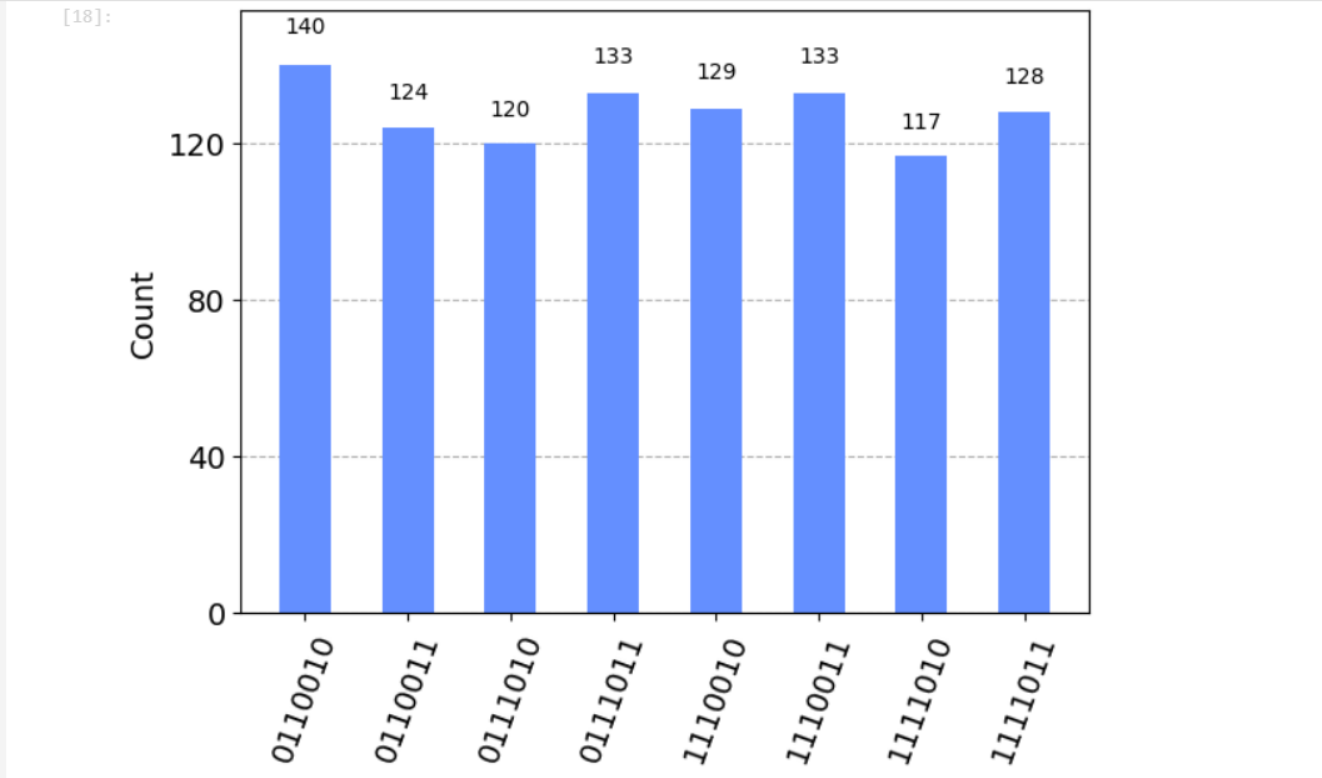}
\caption{Histogram plot of the quantum circuit used in QKD protocol}
\end{figure}
An important point to observe in the plot is that for the common bases i.e., 1, 2, 4, and 5 index no. every key has the same output value. The output corresponding to other index numbers might change but for the common bases, Bob's output always matches with the bits sent by the Alice i.e., 1011. 
\begin{figure}[htp]
\centering
\includegraphics[width=1\textwidth,height=6.5cm]{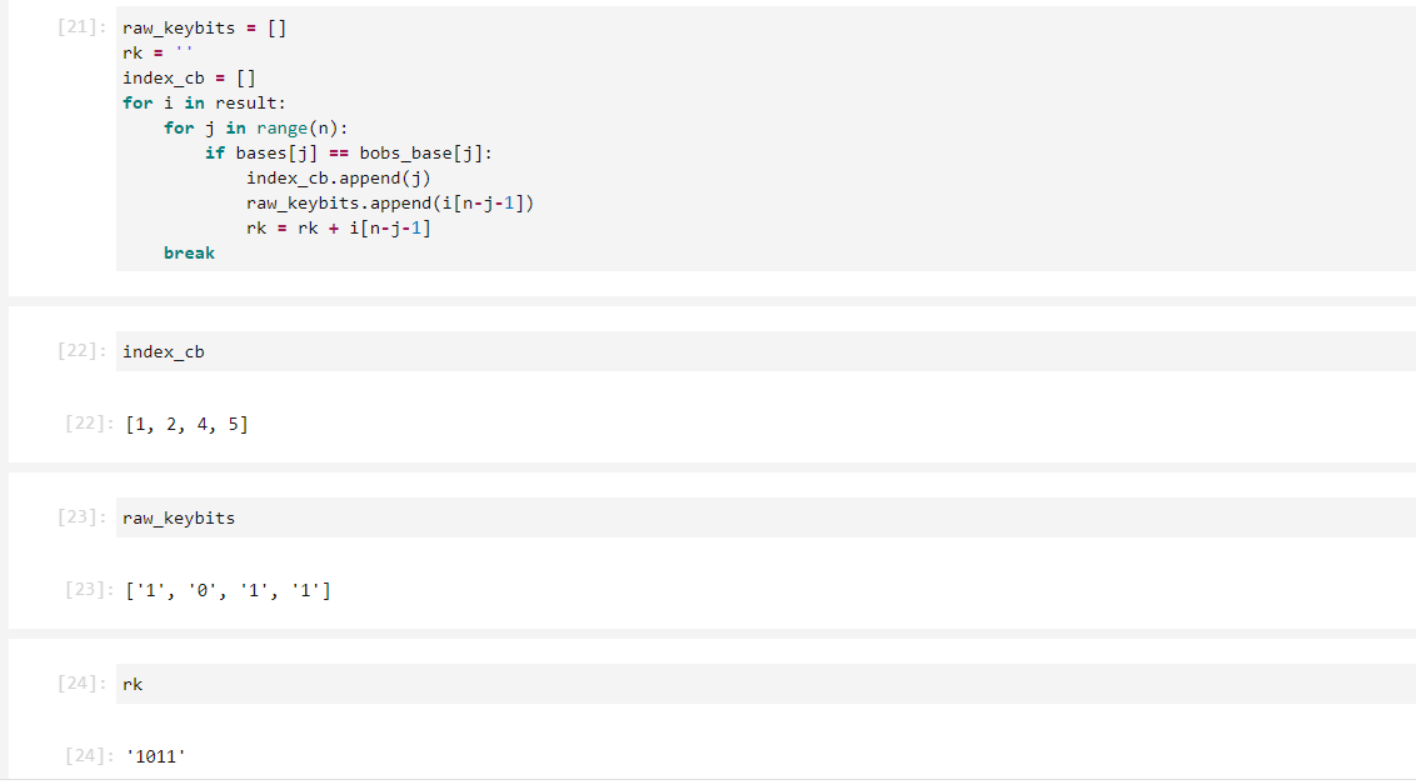}
\caption{Generation of initial key bits from the measurement result}
\end{figure}
This common shared bit string between Alice and Bob is called raw key bits.The code shown above in Figure 9 keeps the record of the common basis, common bases index, and their corresponding measurement bits, which are commonly called raw key bits. 
\newline
\textbf{Raw key bits}
\newline
$[1011]$
\subsection{Eavesdropping, detection of eve, and error correction}
Our simulation of intercept and resend eavesdropping is based on randomly measuring the quantum circuit between the two parties.Our goal is to intercept the quantum circuit before Bob measures it, and to measure it random, and then to transform the state of the qubit based on the measurement result. By doing so, Bob might receive qubits that were not prepared by Alice, but by Eve. We were limited to executing the protocol on the backend simulator since IBM quantum computers only support seven qubits. The generation of sets requires a large number of qubits. A protocol based on 45 qubits has been implemented here.
\begin{figure}[htp]
\centering
\includegraphics[width=1\textwidth,height=7cm]{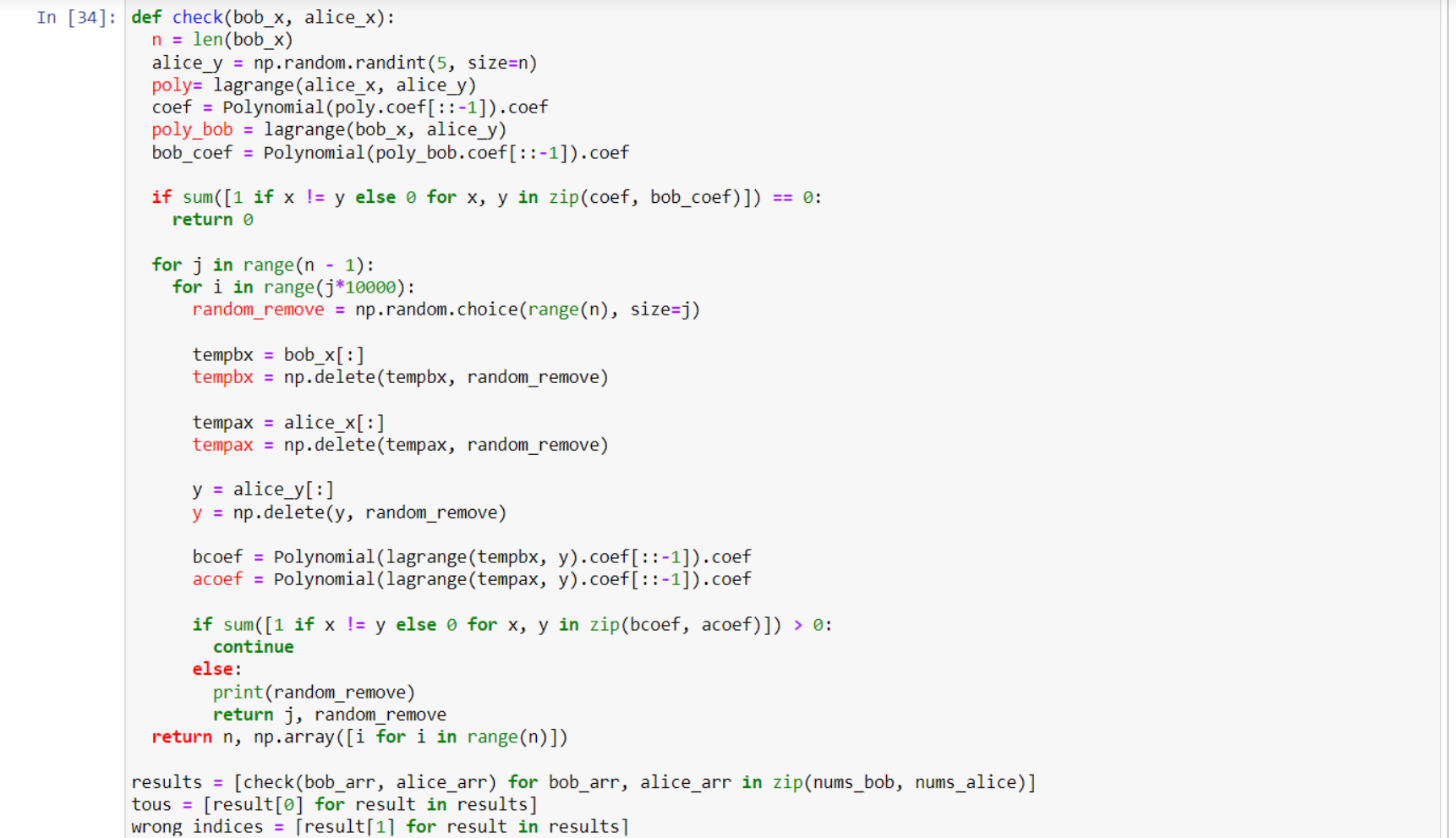}
\caption{Function for detection of Eve and error correction using polynomial interpolation}
\end{figure}In Figure 10 Avg. \% error and bits changed are plotted against $\alpha$. Here a term $\alpha$ has been introduced which is called strength of eavesdropping. It represents the extent of eavesdropping, if $\alpha$ = .1 then it means eavesdropping has been done arbitrarily on 10 \% of the total qubits, 20\% in the case of $\alpha$ = .2, and so on. So let's say $\alpha$ = .2 then in our case(total no. of qubit used is 45) eavesdropping has been done on 9 bits and is represented by bits changed. But \% error has been calculated after checking for the Eve(bits changed in common bases) and error correction using polynomial interpolation. As shown in Fig. 10 the main part of the code whose function is to check for errors using the interpolation technique and calculate the average \% error over the common bases only. Generally, the $\tau$ is considered between 20 - 25 \%. Hence in Fig. 11 for the last 2 cases of  $\alpha$ = .6 and .66, the extent of eavesdropping is so large that $\epsilon$ has become greater than $\tau$ and hence the key generation should be stopped and  process is restarted again.  
\begin{figure}[htp]
\centering
\includegraphics[width=1\textwidth,height=6.5cm]{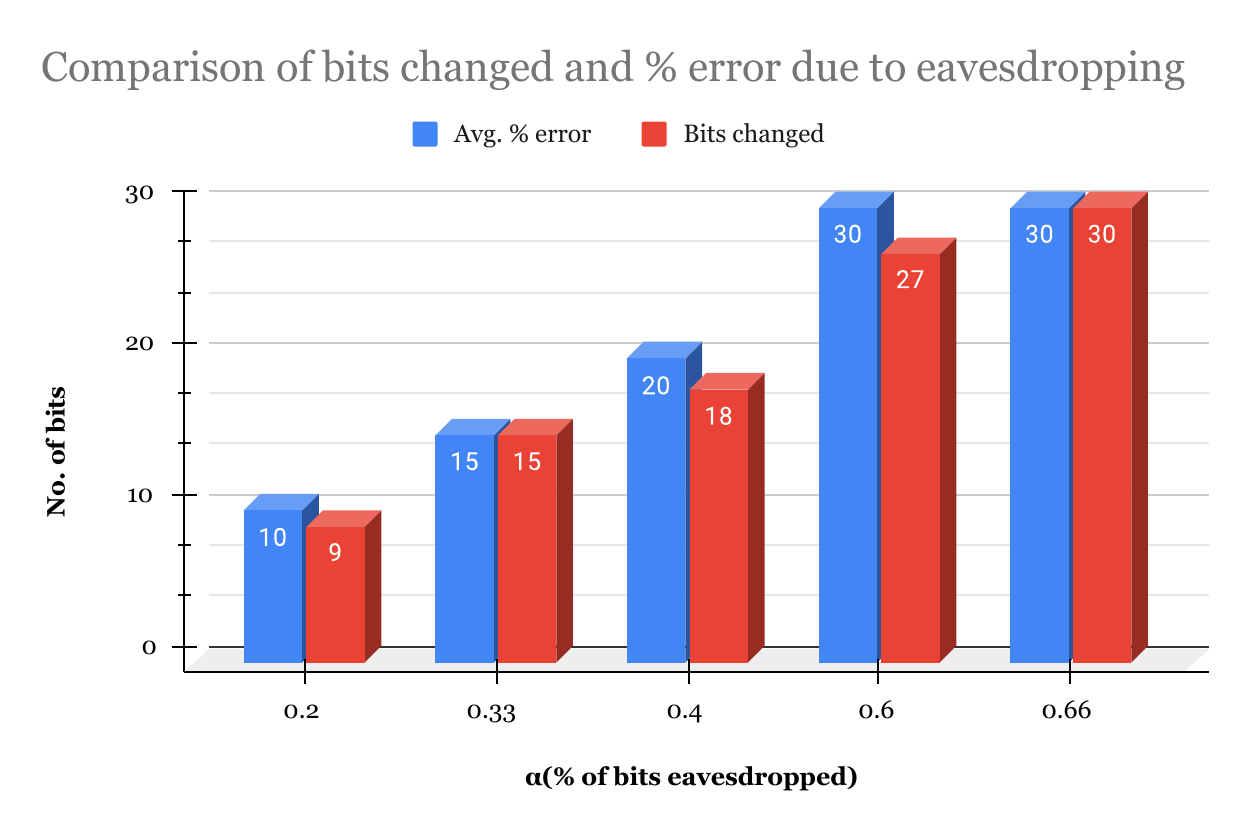}
\caption{ Calculation of \% error against $\alpha$ after error correction using polynomial interpolation}
\end{figure}
\subsection{Introducing noise and calculating success probability of the desired keys}
An unwanted or random interference can disrupt quantum states or disrupt quantum operations. Considering how sensitive qubits are to noise, it's very important to control and fix it for accurate results in quantum computing. Figures 12 and 13 discuss different types of analysis based on noise that can have a large impact on quantum systems. There is a comparison between Pauli gate noise, depolarizing noise, and measurement noise in this study. A quantum calculation error can be caused by Pauli gate noise. At probability p, this executes the gate operation and does nothing at probability 1-p. It is difficult to make accurate calculations when quantum states lose information and do not stay the same due to depolarizing noise. In addition to measurement noise, detectors can be inaccurate, and outside interference can cause measurement bits to flip. Based on the results of Figure 8, several different measurement bit strings are received after the circuit has been run for 1024 shots. It should be noted, however, that for the common bases, all the bits were the same in each sequence, hence they were used as raw key bits.
\begin{figure}[htp]
\centering
\includegraphics[width=0.7\textwidth,height=6.5cm]{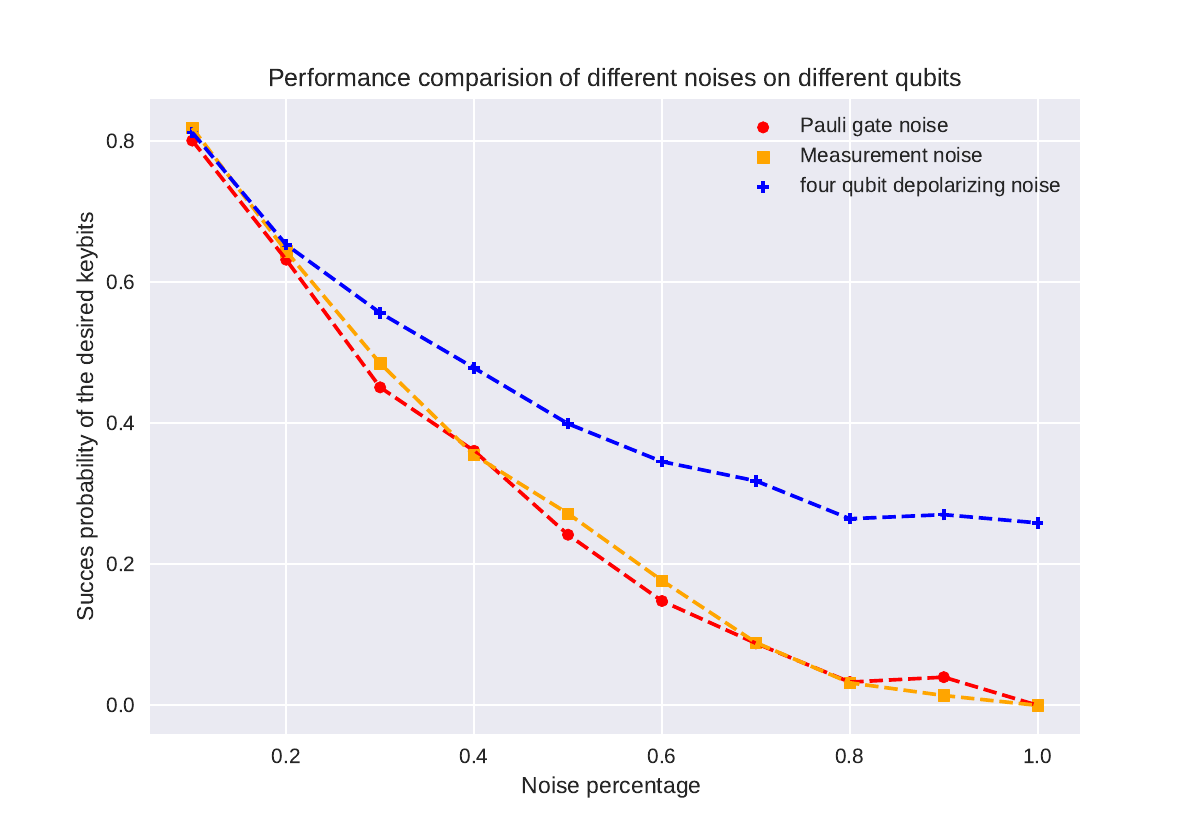}
\caption{Comparison of success probability of desired key bits under the influence of different noises}
\end{figure}
In the case of noise introduced into the quantum circuit, the received sequences may contain different measurements corresponding to common bases. We have plotted the success probability of the desired raw key bits (measurement bits Bob that are the same as bits sent by Alice) against the noise percentage in section 6.1 when executing the seven qubits protocol. A comparison of success probabilities under Pauli gates, measurement errors, and depolarizing errors is shown in figure 12. If the noise percentage was .1, the success probability was 100\%, but with increasing noise percentages, the success probability of the desired keys decreased. When the depolarizing error was present, the success probability decreased, but under 100\% noise, it did not reach zero. With Pauli gate and measurement error, the success probability declined to 0 when noise percentage increased to 1. There is only 1  qubit in this plot, so the depolarizing noise has been applied to it. 
Depolarization refers to the loss of information in each qubit due to the process of depolarization. Therefore, depolarizing noise can be applied to each qubit. The same circuit has been used that was executed in section 6.1. Depolarizing noise has been introduced on different numbers of qubits in Figure 13 in order to compare their impact on success probability. 
\begin{figure}[htp]
\centering
\includegraphics[width=0.6\textwidth,height=6.5cm]{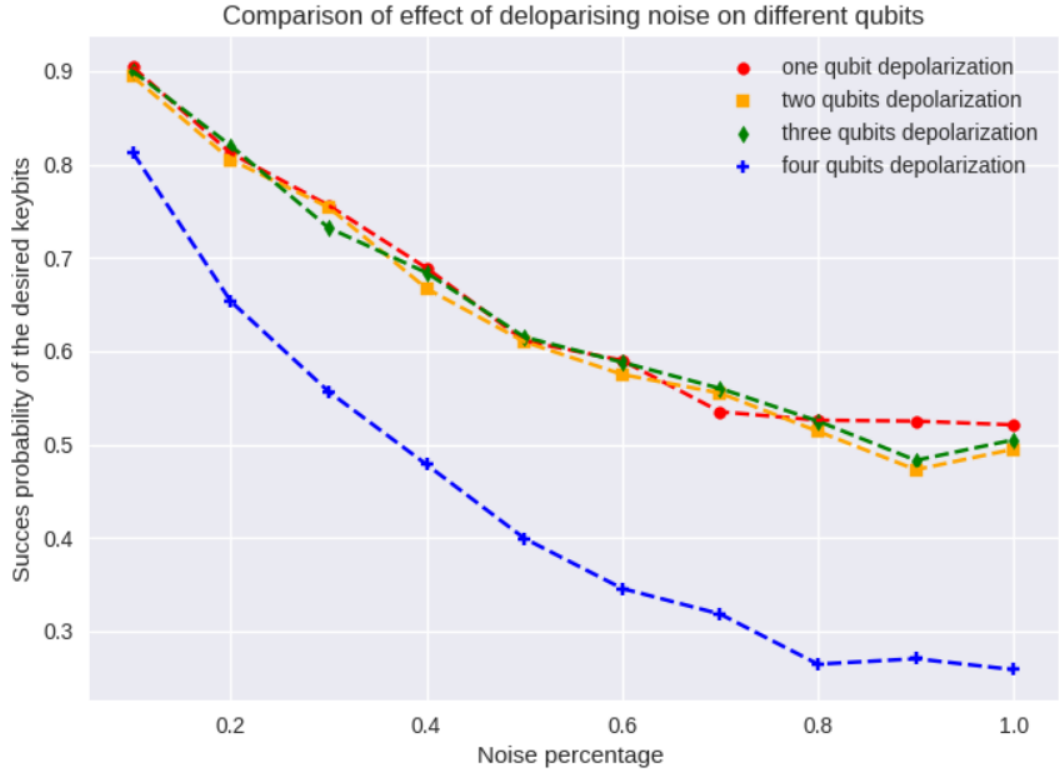}
\caption{Comparison of success probability with depolarizing noise applied to different number of qubits}
\end{figure}The first qubit was introduced with noise, followed by the second, third, and fourth. According to the plot, success probability varies quite similarly with noise percentage for depolarizing errors with one-, two-, and three qubits. As the noise increases, the success probability in these three cases decreases and settles around 50\%, but in case of the four-qubit depolarizing error, the probability drops to 20\%. As a result of the first three cases, we can still deduce the key bits with half the probability, but in the fourth case, it is nearly impossible to generate desired key bits.       

\section{Conclusion}
Quantum cryptography guarantees complete security to both parties. A quantum key distribution protocol is being used to make sure that only authorized parties can access the secret key. Security of the protocol depends on the fact that any attempt to intercept or eavesdrop on communication will be detected. The proposed protocol, however, is defined as a non-leaking encrypted protocol with no information leakage using polynomial interpolation and Eve detection is also discussed as part of the protocol.In the case of quantum circuits where noise is introduced, the received sequences will have different measurements bits corresponding to common bases. Using 7 qubits, the success probability for raw key bits has been plotted against the noise percentage.By analyzing the statistical outcomes of quantum states, the accuracy of quantum states is estimated. As a result performance comparisons of different noises and the effect of depolarizing noise on different qubits are also explained in the work and noise percentage is calculated. In the future, attacks on the discrete system can also be implemented to test and validate the security of the proposed algorithm.
\newline
\newline
\textbf{Credit authorship contribution statement}
\newline
\textbf{Neha Sharma} Conception and design of Study, Acquisition of data, Analysis and or interpretation of data, Writing original draft.
\textbf{Vikas Saxena} has reviewed, edited and verified the work.
\section*{Declarations}{Declaration of Competing Interest}
The authors declare that they have no known competing financial interests or personal relationships that could have appeared to influence the work reported in this paper.

\bibstyle{unsrt}
\bibliography{sn-bibliography}

\end{document}